\documentclass{article}
\usepackage{amsmath}
\usepackage{amsfonts}
\usepackage{amssymb}
\usepackage[usenames]{color}
\usepackage{amsthm}

\def\Th{\Theta}
\def\l{\lambda}
\def\p{\partial}
\def\e{\mathrm{e}}
\newtheorem{theorem}{Theorem}

\theoremstyle{remark}
\newcommand{\wt}{\widetilde}
\newcommand{\be}{\begin{equation}}
\newcommand{\ee}{\end{equation}}
\newcommand{\bea}{\begin{eqnarray}}
\newcommand{\eea}{\end{eqnarray}}
\newcommand{\beaa}{\begin{eqnarray*}}
\newcommand{\eeaa}{\end{eqnarray*}}

\newcommand{\nn}{\nonumber}
\renewcommand{\d}{\mathrm{d}}

\usepackage{authblk}
\begin{document}
%%%%%%%%%%%%%%%%%%%%%%%%%%%%%%%%%%%%%%%%%%%%%
\title
%{Бездисперсионные интегрируемые системы и уравнения
%Богомольного на фоне геометрии Эйнштейна-Вейля}
{On some linear equations associated with dispersionless integrable systems
%\thanks
%{This research was performed in the framework 
%of State Assignment of Ministry of Science and
%Higher Education of the Russian Federation,
%topic 0029-2021-0004 
%(Quantum field theory).}
}
\author{L.V. Bogdanov\thanks{leonid@itp.ac.ru}}
\affil{Landau Institute for Theoretical Physics RAS}
%%%%%%%%%%%%%%%%%%%%%%%%%%%%%%%%%%%%%%%%%%%%%
\date{}
\maketitle
\begin{abstract}
We use a recently proposed scheme
of matrix extension of dispersionless integrable systems
for the Abelian case, in which it leads to linear equations,
connected with the initial dispersionless system. 
In the examples considered,
these equations
can be interpreted in terms of Abelian gauge fields on the
geometric background defined by the dispersionless system.
They are also connected with the linearisation of initial systems.
We construct solutions to these linear equations in terms of
wave functions of the Lax pair for dispersionless system,
which is represented in terms of some vector fields.
\end{abstract}
%%%%%%%%%%%%%%%%%%%%%%%%%%%%%%%%%%%%%%%%%%%%%%%%%%%%%%
%%%%%%%%%%%%%%%%%%%%%%%%%%%%%%%%%%%%%%%%%%%%%%%%%%%

\section{Introduction}
%%%%%%%%%%%%%%%%%%%%%%%%%%%%%%%%%%%%%%%%%%%
Recently we proposed a scheme of matrix extension of the Lax pairs of dispersionless
integrable systems (see \cite{LVB21} and references therein),
which leads to matrix equations on the background of the initial dispersionless systems.
There are important cases in which dispersionless integrable systems describe some geometric structures
(self-dual conformal structures, Einstein-Weyl geometry \cite{DFK15}), and in these cases 
the matrix extension
scheme provides equations for gauge fields 
on the respective geometric background (see \cite{LVB17}, \cite{LVB19}, \cite{LVB20}).
%%%%%%%%%%%%%%%%%%%%%%%%%%%%%%%%%%%%%%%%%%%

In the present work, we would like to develop in more detail some observations
made in the article \cite{LVB21} about the Abelian case
of the matrix extension scheme.
%The article \cite{LVB21} contains some observations about the Abelian case
%of the matrix extension scheme, which we would like to develop in more detail 
%in the present work.
In the Abelian case equations of extension
become linear,
in our examples they can be  represented as an action of 
linear differential operators of the second order (with the coefficients
defined through the solutions of the basic dispersionless integrable system) 
on the scalar
function. But nevertheless 
these equations could
be of interest for several reasons. 
First, in three and
four dimension, where we have an interpretation 
of equations in terms of gauge fields, 
the Abelian case corresponds to electromagnetic fields on
geometric background and could be 
of interest by itself. Second, the arising 
linear operators are connected with the 
linearisation of basic dispersionless equations 
and can be useful for the study of stability of solutions
and singularities of these equations.
For example, for the second heavenly equation
linear operator of Abelian extension is exactly the linearisation
of the equation. And finally, the general solution of extension equations
in the Abelian case
can be found explicitly through the wave functions of the Lax pair.
In the paper \cite{LVB21} it was done using the dressing scheme, but here
we will not use the dressing scheme, constructing the general solution in a rather elementary
way.

Our main examples in this work include the equations of self-dual conformal structure (SDCS)
and the Manakov-Santini system, which describes the Einstein-Weyl structures.
For convenience of the reader, we provide some basic information about the matrix extension scheme
and geometric structures in the Appendices.
%%%%%%%%%%%%%%%%%%%%%%%%%%%%%%%%%%%%%%%%%%%%
\section{Abelian extension of SDCS equations}
%%%%%%%%%%%%%%%%%%%%%%%%%%%%%%%%%%%%%%%%%%%%
Let us consider the Lax pair \cite{BDM07}
\bea
\begin{aligned}
X_1&=\p_z-\lambda\p_x
+F_x\p_x+G_x\p_y+ f_x\p_\lambda,
\\
X_2&=\p_w- \lambda\p_y + F_y\p_x+G_y\p_y+f_y\p_\lambda.
\label{LaxSDCS}
\end{aligned}
\eea
Commutation relations for this Lax pair give a
coupled system of three second-order PDEs for the functions $F$, $G$, $f$
\bea
\left\{
\begin{aligned}
Q(F)&=f_y,
\\
Q(G)&=-f_x,
\\
Q(f)&=0,
\end{aligned}
\right.
\label{SDCS3}
\eea
where linear second order differential operator $Q$ is given by
\bea
\begin{aligned}
Q&=(\p_w+F_y\p_x+G_y\p_y)\p_x+(\p_z+F_x\p_x+G_x\p_y)\p_y
\\
&=\p_w\p_x-\p_z\p_y+F_y{\p_x}^2-G_x{\p_y}^2-
(F_x-G_y)\p_x\p_y.
\end{aligned}
\label{Q}
\eea
System (\ref{SDCS3}) can be rewritten in the form of a
coupled system of third order PDEs for the functions $F$, $G$
\bea
\left\{
\begin{aligned}
&
\p_x(Q(F))+\p_y(Q(G))=0,
\\
&
(\p_w+F_y\p_x+G_y\p_y)Q(G)+(\p_z+F_x\p_x+G_x\p_y)Q(F)=0,
\label{sd_3rd}
\end{aligned}
\right.
\label{SDCS2}
\eea
in this form it was introduced in \cite{DFK15} 
in connection with the self-dual conformal structures (see Appendix 2).

Scalar extension of the Lax pair (\ref{LaxSDCS})
(see Appendix 1)
\beaa
\begin{aligned}
%\label{new_lax}
\nabla_{X_1}&=\p_z-\lambda\p_x
+F_x\p_x+G_x\p_y+ f_x\p_\lambda + a_1,
\\
\nabla_{X_2}&=\p_w- \lambda\p_y + F_y\p_x+G_y\p_y+f_y\p_\lambda + a_2
\end{aligned}
\eeaa 
generates a linear equation for the potential $\phi$, $a_1=\p_x\phi$, $a_2=\p_y\phi$,
\bea
%&&
Q\phi:=(\p_w\p_x-\p_z\p_y+F_y{\p_x}^2-G_x{\p_y}^2-
(F_x-G_y)\p_x\p_y)\phi=0.
\label{linSD}
\eea
%\\
%&&
%\quad
%Q=\p_w\p_x-\p_z\p_y+F_y{\p_x}^2-G_x{\p_y}^2-(F_x-G_y)\p_x\p_y
%\eeaa
%%%%%%%%%%%%%%%%%%%%%%%%%%%%%%%%%%%%%%%%%%%%%
\subsection*{Solution through the wave functions.}
%%%%%%%%%%%%%%%%%%%%%%%%%%%%%%%%%%%%%%%%%%%%%%%
In \cite{LVB21} we constructed a general solution to linear equation (\ref{linSD})
using a dressing scheme.
It is easy to obtain this formula directly using  the dispersionless Lax pair. Indeed,
cross-differentiating over $y,x$ linear equations
\bea
\begin{aligned}
\label{LinLaxSD}
(\p_z-\lambda\p_x
+F_x\p_x+G_x\p_y+ f_x\p_\lambda)\Psi&=0,
\\
(\p_w- \lambda\p_y + F_y\p_x+G_y\p_y+f_y\p_\lambda)\Psi&=0,
\end{aligned}
\eea
we get the relation
\bea
\left(
(\partial_w+F_y \partial_x + G_y\p_y )\partial_x
-
(\partial_z+ F_x \partial_x + G_x\p_y)\partial_y 
\right)\Psi
=\p_\l(f_x\p_y - f_y\p_x ) \Psi.
\label{cross}
\eea
Integration of the r.h.s. with respect to $\lambda$ over a closed contour gives zero, thus
$Q\oint\Psi d\lambda=0$, and
\bea
\phi=
\frac{1}{2\pi i}\oint\Psi d\lambda
\label{solSD}
\eea 
gives a solution to linear equation (\ref{linSD}) for an arbitrary wave function (analytic in the neihborhood
of the contour or given in terms of formal  Laurent series). Linear equations
(\ref{LinLaxSD}) possess three basic wave functions $\Psi^0$,  $\Psi^1$,   $\Psi^2$ 
\cite{BDM07} and a general
wave function is represented as 
\beaa
\Psi=f(\Psi^0, \Psi^1, \Psi^2) ,
\eeaa
where $f$ is an arbitrary analytic function. Thus solution (\ref{solSD}) possesses a functional freedom
of a function of three variables, corresponding to a general solution of linear equation (\ref{linSD}).

For trivial background
\beaa
&&
\left(
\partial_w\p_x
-\partial_z\p_y 
\right)\phi=0,
%\label{Qq}
%\eeaa
\\
%\beaa
&&
\phi=
\frac{1}{2\pi i}\oint 
f(\l, \l z +x,\l w + y) d\lambda.
%\label{Psi0}
\eeaa
This formula is easily recognised as a version
of Penrose formula 
for solutions of the wave equation
written for the case of neutral signature.

Considering  the linearization of SDCS system (\ref{SDCS3}) (or (\ref{SDCS2})), we come to the observation
that operator $Q$ is contained as a factor in the principal part  of linearised equations.
%%%%%%%%%%%%%%%%%%%%%%%%%%%%%%%%%%%%%%%%%%%%
\subsection*{Reductions of the SDCS system}
%%%%%%%%%%%%%%%%%%%%%%%%%%%%%%%%%%%%%%%%%%%%%
First, let us consider the volume-preserving reduction, that leads to the Dunajski system
\cite{Dun02}.
This reduction is connected with zero divergence vector fields
(\ref{LaxSDCS}). In this case functions $F$, $G$ can be defined through the potential $\Theta$,
$F=\Theta_y$, $G=-\Theta_x$, and system (\ref{SDCS3}) takes the form
\bea
\begin{aligned}
\label{Dun}
&\Th_{wx}+\Th_{zy}+\Th_{xx}\Th_{yy}-\Th_{xy}^2=f,
\\
&
Q f=0,
\end{aligned}
\eea
where the operator $Q$ is expressed in terms of potential $\Theta$ as
\beaa
Q=
%\left(
\partial_w\p_x
+\partial_z\p_y 
+\Theta_{yy}\p_x\p_x
+\Theta_{xx}\p_y\p_y 
-2\Theta_{xy}\p_x\p_y, 
%\right)
%\label{QHeav}
\eeaa
This operator defines an Abelian extension of the Dunajsky system,
it also represents a bivector defining a conformal structure. 
System (\ref{Dun}) can be written as one fourth-order equation
\beaa
Q (\Th_{wx}+\Th_{zy}+\Th_{xx}\Th_{yy}-\Th_{xy}^2)=0
\eeaa

Another standard reduction 
of SDCS system (\ref{SDCS3})
is a linearly degenerate case, 
which corresponds to hyper CR (Cauchy-Riemann) equations.
In this case vector fields (\ref{LaxSDCS}) do not contain a derivative over spectral variable,
$f=0$, $\Psi^0=\lambda$ is a wave function of linear operators,
and system (\ref{SDCS3}) reads
\bea
\left\{
\begin{aligned}
Q(F)&=0,
\\
Q(G)&=0.
\end{aligned}
\right.
\label{SDCS03}
\eea
Operator $Q$ here is of the same form as in general SDSC case, it also defines
an abelian extension. However, due to the
reduction, some new special features of solutions of this operator arise. Indeed, 
for the reduced Lax pair 
\beaa
\begin{aligned}
%\label{LinLaxSD}
(\p_z-\lambda\p_x
+F_x\p_x+G_x\p_y)\Psi&=0,
\\
(\p_w- \lambda\p_y + F_y\p_x+G_y\p_y)\Psi&=0,
\end{aligned}
\eeaa
and instead of relation (\ref{cross}) we now have
\bea
Q\Psi=\left(
(\partial_w+F_y \partial_x + G_y\p_y )\partial_x
-
(\partial_z+ F_x \partial_x + G_x\p_y)\partial_y 
\right)\Psi
=0.
\label{cross1}
\eea
Thus  for linearly degenerate case {\em an arbitrary wave function of the reduced  Lax pair satisfies 
equation (\ref{linSD})}! The reduction also leads to a recursion for
solutions of equation (\ref{cross1}), defined by the relations
\bea
\begin{aligned}
\label{recursion}
\p_x\phi' &=(\p_z
+F_x\p_x+G_x\p_y)\phi,
\\
\p_y\phi' &=(\p_w+ F_y\p_x+G_y\p_y)\phi.
\end{aligned}
\eea
The compatibility condition for these relations coincide with equation  (\ref{cross1}) for $\phi$,
and cross-action of linear operators of the r.h.s. gives (modulo equation (\ref{SDCS03}))
equation  (\ref{cross1}) for $\phi'$.

Similar observations for operators of linearisation of linearly degenerate equations 
were done in the work of Segyeyev \cite{Serg17}.

Finally, applying both volume-preserving and linearly degenerate case reductions to SDCS system
(\ref{SDCS3}), we obtain the famous Pleba\'nski second heavenly equation
\bea
&&
\Th_{wx}+\Th_{zy}+\Th_{xx}\Th_{yy}-\Th_{xy}^2=0.
\label{Heav}
\eea
The operator of Abelian extension is of the same form as for the Dunajski system,
\beaa
Q=
%\left(
\partial_w\p_x
+\partial_z\p_y 
+\Theta_{yy}\p_x\p_x
+\Theta_{xx}\p_y\p_y 
-2\Theta_{xy}\p_x\p_y, 
%\right)
%\label{QHeav}
\eeaa
it coincides with the linearisation operator for the heavenly equation (\ref{Heav}). 
Solutions to the equation $Q\phi=0$ are given by arbitrary wave functions  of the Lax pair
for the heavenly equation, $\phi=\Psi$,
\beaa
%\label{new_lax}
(\p_z-\lambda\p_x
+\Theta_{xy}\p_x-\Theta_{xx}\p_y)\Psi=0,
\\
(\p_w- \lambda\p_y + \Theta_{yy}\p_x - \Theta_{xy}\p_y)\Psi=0.
\eeaa
Recursion relations for coefficients of expansion of wave functions into the powers of the spectral
variable lead to the  recursion for solutions of the linearized second heavenly equation
$Q\phi=0$,
\beaa
%\label{new_lax}
&&
\p_x \phi'=
(\p_z
+\Theta_{xy}\p_x-\Theta_{xx}\p_y)\phi,
\\
&&
\p_y \phi'=
(\p_w + \Theta_{yy}\p_x - \Theta_{xy}\p_y)\phi.
\eeaa
This type of recursion was introduced in \cite{Serg17}.
%%%%%%%%%%%%%%%%%%%%%%%%%%%%%%%%%%%%%%%%%%%%%
\section{Abelian extension of the Manakov-Santini system}
%%%%%%%%%%%%%%%%%%%%%%%%%%%%%%%%%%%%%%%%%%%
%\subsection*{The basic MS system}
The Manakov-Santini system \cite{MS07} is a two-component integrable 
generalisation of the
dispersionless Kadomtsev-Petviashvili (dKP) equation,
\begin{equation}
\begin{aligned}
u_{xt} &= u_{yy}+(uu_x)_x+v_xu_{xy}-u_{xx}v_y,
\\
v_{xt} &= v_{yy}+uv_{xx}+v_xv_{xy}-v_{xx}v_y
\end{aligned}
\label{MSeq}
\end{equation}
It correspons to arbitrary vector fields in  the Lax pair, instead of Hamiltonian vector fields
for the dKP equation,
\begin{equation}
\begin{aligned}
X_1&=\partial_y-(\l-v_{x})\partial_x + 
u_{x}\partial_\lambda,
\\
X_2&=\partial_t-(\l^2-v_{x}\l+u -v_{y})\partial_x
+(u_{x}\l+u_{y})\partial_\lambda
\end{aligned}
\label{MSLax}
\end{equation}
For $v=0$  this system reduces to the
dKP (or Khohlov-Zabolotskaya) equation
\beaa
u_{xt} = u_{yy}+(uu_x)_x,
%\label{dKP-eq}
\eeaa
reduction $u=0$ (linearly degenerate case) gives the equation 
(Mikhalev \cite{Mikh92}, Pavlov \cite{Pavlov03})
\begin{equation*}
 v_{xt} = v_{yy}+v_xv_{xy} - v_{xx}v_y.
%\label{Pavlov}
\end{equation*}
%%%%%%%%%%%%%%%%%%%%%%%%%%%%%%%%%%%%%%%%%%%%%%
%%%%%%%%%%%%%%%%%%%%%%%%%%%%%%%%%%%%%%%%%%
Abelian extension of the Lax pair (\ref{MSLax})  (see also Appendix 1)
\bea
\begin{aligned}
\nabla_{X_1}&=
\partial_y-(\lambda-v_{x})\partial_x + u_{x}\partial_\lambda + \kappa_x,
\\
\nabla_{X_2}&=\partial_t-(\lambda^2-v_{x}\lambda + u -v_{y})\partial_x
+(u_{x}\lambda+u_{y})\partial_\lambda +\lambda \kappa_x + \kappa_y,
\label{MSLaxE}
\end{aligned}
\eea
leads to linear equation for the scalar function $\kappa$
\beaa
Q\kappa:=(\p_t\p_x-\p_y\p_y - (u-v_y)\p_x\p_x -  v_x\p_x\p_y- u_x\p_x)\kappa=0.
\eeaa
Any wave function $\Psi(\lambda)$ of linear operators (\ref{MSLax}) 
given on the contour
defines a solution to this equation by the formula analogous to (\ref{solSD}).

To derive this formula directlly from the Lax pair (\ref{MSLax}), we rewrite equations
for the wave functions in the form
\beaa
\begin{split}
&(\partial_y-\l\p_x +v_{x}\partial_x + 
u_{x}\partial_\lambda)\Psi=0,
\\
&(\partial_t- \l \p_y
-(u -v_{y})\partial_x
+u_{y}\partial_\lambda)\Psi=0.
\end{split}
%\label{MSLax}
\eeaa
Cross-differentiating by respectively $y$ and $x$ and taking the difference, we  get
\bea
Q\Psi= \p_\lambda (u_x\p_y-  u_y\p_x) \Psi ,
\label{cross2}
\eea
then integration over the contour cancels the r.h.s., and we obtain the formula 
\bea
\kappa=
\frac{1}{2\pi i}\oint 
\Psi(\mu) d\mu.
\label{Psi-kappa}
\eea
In terms of the basic wave functions of the Lax pair (\ref{MSLax})
$
\Psi=F(\Psi^0, \Psi^1) ,
$
and we have a solution with functional freedom of a function of two variables.
%%%%%%%%%%%%%%%%%%%%%%%%%%%%%%%%%%%%%%%%%%%%%%%%%%%%%%%%%
%%%%%%%%%%%%%%%%%%%%%%%%%%%%%%%%%%%%%%%%%%%%%
\subsection*{The dKP equation case}
%%%%%%%%%%%%%%%%%%%%%%%%%%%%%%%%%%%
The case of the dKP equation corresponds to Hamiltonian vector fields 
in the Lax pair (\ref{MSLax}), leading to  $v=0$ and 
\bea
u_{xt} = u_{yy}+(uu_x)_x.
\label{dKP}
\eea
Commutation relations for the extended Lax pair
(reduced extended Lax pair (\ref{MSLaxE}))
\beaa
&&
\nabla_{X_1}=
\partial_y-\lambda\partial_x + u_{x}\partial_\lambda + \kappa_x,
\nn\\
&&
\nabla_{X_2}=\partial_t-(\lambda^2+ u )\partial_x
+(u_{x}\lambda+u_{y})\partial_\lambda +\lambda \kappa_x + \kappa_y
%\label{MSLax-ext}
\eeaa
imply the dKP equation (\ref{dKP})
and linear equation for $\kappa$,
$$
Q\kappa:=(\p_t\p_x - \p_y\p_y - u \p_x\p_x - u_x\p_x)\kappa=0.
$$
The operator $Q$ doesn't coincide with the linearisation operator for dKP equation (\ref{dKP}),
which reads
$$
P=\p_t\p_x - \p_y\p_y - u \p_{xx} - 2u_x\p_x- u_{xx}
$$
However, the two operators are connected via a simple identity
$
\p_x Q = P\p_x,  
$
which implies that $Q\kappa=0\Rightarrow P\p_x \kappa=0$.
In other  words, the operator $Q$ corresponds to the linearisation
of the potential dKP equation for the function $w$, $u=w_x$.
For $\kappa$ we have a formula (\ref{Psi-kappa}),
\beaa
\kappa=
\frac{1}{2\pi i}\oint 
F(\Psi^0,\Psi^1) d\mu,
%\label{Psi0}
\eeaa
it  gives a symmetry of potential dKP equation.
The symmetry for the dKP equation (solution of linearised equation) is defined by $\p_x \kappa$.
This is a rather familiar formula for the symmetries of the dKP equation,
in standard dKP hierarchy notations $\Psi^0=L$ (Lax-Sato function), $\Psi^1=M$ (Orlov function).
%%%%%%%%%%%%%%%%%%%%%%%%%%%%%%%%%%%%%%%%%%
\subsection*{Mikhalev-Pavlov equation}
%%%%%%%%%%%%%%%%%%%%%%%%%%%%%%%%%%%%%%%%%%%
Considering the linearly degenerate case of the MS system, for which $u=0$, we get
the equation
\bea
v_{xt} &= v_{yy} + v_x v_{xy}-v_{xx} v_y
\label{eqPavlov}
\eea
with the Lax pair
\bea
\begin{split}
{X_1}&=\partial_y-\l\p_x + v_{x}\partial_x ,
\\
{X_2}&=\partial_t-(\l^2-v_{x}\l -v_{y})\partial_x .
%=\partial_t - \l \p_y + v_{y}\partial_x +\lambda \kappa_x + \kappa_y
\end{split}
\label{linMP}
\eea
The Abelian extension of the Lax pair
\bea
\begin{split}
\nabla_{X_1}&=\partial_y-\l\p_x + v_{x}\partial_x +\kappa_x ,
\\
\nabla_{X_2}&=\partial_t-(\l^2-v_{x}\l -v_{y})\partial_x +\lambda \kappa_x + \kappa_y
%=\partial_t - \l \p_y + v_{y}\partial_x +\lambda \kappa_x + \kappa_y
\end{split}
\label{linMPE}
\eea
implies linear equation 
\bea
Q\kappa:=(\p_t\p_x-\p_y\p_y + v_y\p_x\p_x - v_x\p_x\p_y)\kappa=0. 
\label{AeMP}
\eea
Similar to linearly degenerate case of the SDCS equations (\ref{cross1}),
{\em any wave function of linear operators (\ref{linMP}) $\Psi(\l)$,
$X_1\Psi(\l)=0$, $X_2\Psi(\l)=0$, satisfies this  linear equation}. 
Indeed, the r.h.s. of the formula (\ref{cross2}) in the linearly degenerate case
is equal to zero,
then $Q\Psi(\l)=0$. In terms of the basic wave functions of linear operators (\ref{linMP}),
$\Psi=F(\l,\Psi^1)$.

Linear operator $Q$ in this case does not coincide with the
linearisation operator for equation (\ref{eqPavlov}), which is
\begin{multline}
P=\p_t\p_x-\p_y\p_y + v_y\p_x\p_x - v_x\p_x\p_y +v_{xx}\p_y - v_{xy}\p_x
\\
=Q+v_{xx}\p_y - v_{xy}\p_x.
\label{symMP}
\end{multline}
In the work \cite{Serg17} it was demonstrated  that the linearised equation
is satisfied by the function $\Psi_x^{-1}$,
$P\Psi_x^{-1}=0$, where $\Psi(\l)$ is a wave function of the Lax pair (\ref{linMP}).
The recursion for the linearised equation was also constructed.

To derive linearisation  operator and solutions for it in terms of the Lax pair,
we will use a parametric deformation of the Lax pair described in Appendix 3,
\bea
\begin{split}
X_{1\alpha} &=\partial_y-\l\p_x + v_{x}\partial_x + \alpha  v_{xx} ,
\\
X_{2\alpha} &=\partial_t - \l \p_y + v_{y}\partial_x + \alpha  v_{xy}.
\end{split}
\label{defMP}
\eea
This deformation takes us out of the class of vector fields, however, the compatibility 
conditions
remain the same.
For $\alpha=0$ this is a standard Lax pair in terms of vector fields, 
and $\alpha=1$ corresponds to the formally adjoint
Lax pair. The deformed Lax pair implies a special solution
for Abelian extension equation (\ref{AeMP}) $\kappa=v_x$,
that is easily checked by differentiating Mikhalev-Pavlov equation
(\ref{eqPavlov}).
A general wave function of deformed Lax pair (\ref{defMP}) 
in terms of the basic functions
of the Lax pair (\ref{linMP}) reads (see Appendix 3)
\beaa
\wt\Psi_\alpha= (\Psi^1_x)^\alpha F(\l,\Psi^1).
\eeaa
Instead of formula (\ref{cross2}) we obtain
\beaa
\partial_y(\partial_y+v_{x}\partial_x+\alpha  v_{xx})\wt\Psi_\alpha(\l)
=\partial_x(\partial_t + v_{y}\partial_x  + \alpha  v_{xy})\wt\Psi_\alpha(\l)
,
%\label{cross}
\eeaa
thus 
\beaa
&&
Q_\alpha\wt\Psi_\alpha:= \left(\p_t\p_x-\p_y\p_y + v_y\p_x\p_x - v_x\p_x\p_y +\alpha  (v_{xy}\p_x- v_{xx}\p_y)
\right)\wt\Psi_\alpha=0.
%\\&&
%\wt\Psi_\alpha= (\Psi^1_x)^\alpha F(\l,\Psi^1),
\eeaa
For solutions of linear equation $Q_\alpha \phi=0$ we have a recursion 
\beaa
&&
\phi'_x=(\partial_y+ v_{x}\partial_x + \alpha  v_{xx})\phi,
\\
&&
\phi'_y=(\partial_t + v_{y}\partial_x + \alpha  v_{xy})\phi.
\eeaa
The compatibility condition given by cross-differentiation over $y,x$ is the equation 
$Q_\alpha \phi=0$, while cross-action of linear operators of the r.h.s. leads 
(modulo equation (\ref{eqPavlov})) to the equation
$Q_\alpha \phi'=0$.
The case of linearisation operator $P$ corresponds to $\alpha=-1$, $P=Q_{-1}$. Symmetries 
for the Mikhalev-Pavlov equation are given by the expression
\beaa
\wt\Psi_{-1}= (\Psi^1_x)^{-1} F(\l,\Psi^1),
\eeaa 
and recursion for the symmetries is
\beaa
&&
\phi'_x=(\partial_y+ v_{x}\partial_x -  v_{xx})\phi,
\\
&&
\phi'_y=(\partial_t + v_{y}\partial_x -  v_{xy})\phi.
\eeaa
%%%%%%%%%%%%%%%%%%%%%%%%%%%%%%%%%%%%%%%%%%%
\subsection*{The interpolating reduction}
%%%%%%%%%%%%%%%%%%%%%%%%%%%%%%%%%%%%%%%%%%%
Let us consider also the interpolating reduction of the Manakov-Santini system (\ref{MSeq}), which 
is defined by the condition
$$
a u=v_x,
$$
where $a$ is a parameter (see \cite{Dun08}, \cite{LVB10}). Under this condition, 
the Manakov-Santini system
can be written as one equation for the function $v$,
\begin{equation}
%\begin{aligned}
%u_{xt} &= u_{yy}+(uu_x)_x+v_xu_{xy}-u_{xx}v_y,
%\\
v_{xt} = v_{yy}+a^{-1}v_x v_{xx}+v_xv_{xy}-v_{xx}v_y.
%\end{aligned}
\label{MSint}
\end{equation}
A remarkable property of this equation, justifying the name `interpolating' \cite{Dun08},
is that its limit for $a\rightarrow 0$ leads to the
dKP equation, while for  $a\rightarrow \infty$ it gives the Mikhalev-Pavlov equation.
The linear equation of Abelian extension in this case reads
\beaa
Q\kappa:=(\p_t\p_x-\p_y\p_y - (a^{-1}v_x-v_y)\p_x\p_x -  v_x\p_x\p_y- 
a^{-1}v_{xx}\p_x)\kappa=0,
\eeaa
and its solution can be obtained using an arbitrary wave function of Lax operators 
(\ref{MSLax}) (taking into account reduction condition) by formula
(\ref{Psi-kappa}).

Similar to the case of the Mikhalev-Pavlov equation, linear operator $Q$ does not coincide
with the linearisation operator for equation (\ref{MSint}), which reads
$$
P=Q+v_{xx}\p_y - v_{xy}\p_x.
$$
To find solutions of equation $P\phi=0$
through the wave functions of Lax operators, we use the same trick as 
for the Mikhalev-Pavlov equation case. 
Introducing a parametric deformation of the Lax pair described in Appendix 3,
we arrive to the following formula for solutions of linearised interpolating equation
$P\phi=0$
(symmetries of interpolating equation)
\beaa
\phi=
\frac{1}{2\pi i}\oint \e^{a(\mu - \Psi^0 )}
\Psi(\mu) d\mu.
%\label{Psi-kappa}
\eeaa
where $\Psi$ is an arbitrary wave function $\Psi=F(\Psi^0,\Psi^1)$, and $\Psi^0$ is a basic 
wave function with the expansion $\lambda + u \lambda^{-1} + \dots$ (corresponds
to Lax-Sato function $L$ in standard dKP notations).
%%%%%%%%%%%%%%%%%%%%%%%%%%%%%%%%%%%%%%%%%%%%
\section*{Appendix 1}
%%%%%%%%%%%%%%%%%%%%%%%%%%%%%%%%%%%%%%%%%%%%
\subsection*{Matrix and Abelian extension of 
multidimensional dispersionless integrable systems}
%%%%%%%%%%%%%%%%%%%%%%%%%%%%%%%%%%%%%%%%%%%
We will give a brief description of the
scheme of matrix extension of the Lax pairs of dispersionless
integrable systems (see \cite{LVB21} and references therein).
Multidimensional dispersionless integrable systems are associated with Lax pairs in terms of vector
fields depending on a spectral parameter. We will consider Lax pairs of the type
\bea
&&
[X_1, X_2]=0,
\\&&
X_1=\p_{t_1}
+\sum_{i=1}^N F_i\p_{x_i} + F_0\p_\lambda,\quad
X_2=\p_{t_2}+ \sum_{i=1}^N G_i\p_{x_i}+G_0\p_\lambda.
\label{Lax0}
\eea
$\lambda$ - `spectral parameter', functions $F_k$, 
$G_k$ are holomorphic in $\lambda$ (polynomials, Laurent polynomials) and depend on the variables $t_1$, $t_2$, $x_n$.
The class of equations corresponding to such Lax pairs includes 
dispersionless limits of integrable equations 
(dKP, dispersionless 2DTL hierarchy), Pleba\'nski 
heavenly equations and their generalizations, hyper-K\"ahler hierarchies.
A scheme of matrix extension 
%(\ref{Lax}) (see \cite{LVB17,LVB19,LVB20})
leads to gauge covariant Lax pairs of the type
\bea
\label{Laxext}
\nabla_{X_1}=X_1 + A_1,
\quad
\nabla_{X_2}=X_2 + A_2,
\eea
$A_1$, $A_2$ are matrix-valued functions of 
space-time variables holomorphic in $\lambda$ 
(polynomials, Laurent polynomials). 
Lax pairs of this structure were already present 
in the seminal work of Zakharov and Shabat (1979).
The commutator of two covariant (extended) vector fields contains
vector field part and matrix (Lie algebraic) part,
\beaa
[\nabla_{X_1}, \nabla_{X_2}]=
[X_1, X_2]+ X_1 A_2 - X_2 A_1 +[A_1,A_2]
\eeaa
Vector fields term of compatibility conditions gives the basic dispersionless system,
\beaa
[X_1, X_2]=0,
\eeaa
while the matrix term provides matrix equations on the dispersionless background
\bea
X_1 A_2 - X_2 A_1 +[A_1,A_2]=0.
\label{background}
\eea
In several important examples the basic system corresponds to some geometric object,
and equations (\ref{background}) are connected with gauge fields on the geometric background
(see Appendix 2).

For Abelian gauge fields,
\bea
\label{new_lax0}
\nabla_{X_1}=X_1 + a_1,
\quad
\nabla_{X_2}=X_2 + a_2,
\eea
where $a_1$, $a_2$ are scalar functions, equations  (\ref{background}) become linear,
\bea
X_1 a_2 - X_2 a_1=0.
\label{backgroundA}
\eea
These equations are the main object of the study in the present work. 
%%%%%%%%%%%%%%%%%%%%%%%%%%%%%%%%%%
\paragraph*{Extension of the SDCS equations}
Matrix extension of the Lax pair (\ref{LaxSDCS}) reads
\beaa
\begin{aligned}
%\label{new_lax}
\nabla_{X_1}&=\p_z-\lambda\p_x
+F_x\p_x+G_x\p_y+ f_x\p_\lambda + \Phi_x,
\\
\nabla_{X_2}&=\p_w- \lambda\p_y + F_y\p_x+G_y\p_y+f_y\p_\lambda + \Phi_y,
\end{aligned}
\eeaa 
it generates an equation for the matrix potential $\Phi$ on the background of the
SDCS equations, 
\bea
&&
Q\Phi=[\Phi_x,\Phi_y],
\nn\\
&&
Q=\p_w\p_x-\p_z\p_y+F_y{\p_x}^2-G_x{\p_y}^2-
(F_x-G_y)\p_x\p_y.
\label{linSDm}
\eea
In the Abelian case we have a linear equation for the scalar potential 
$Q\phi=0$.
%%%%%%%%%%%%%%%%%%%%%%%%%%%%%%%%%%%%%%%
\paragraph*{The MS system - extension} 
%%%%%%%%%%%%%%%%%%%%%%%%%%%%%%%%%%%%%%%
Matrix extension of the Lax pair for the MS system reads
\bea
\begin{aligned}
\nabla_{X_1}&=
\partial_y-(\lambda-v_{x})\partial_x + u_{x}\partial_\lambda + A,
\\
\nabla_{X_2}&=\partial_t-(\lambda^2-v_{x}\lambda + u -v_{y})\partial_x
+(u_{x}\lambda+u_{y})\partial_\lambda +\lambda A + B,
\label{MSLax-ext}
\end{aligned}
%\label{MSLax-ext}
\eea
where $A$, $B$ are matrix-valued functions. Vector field
part of commutation relations gives the Manakov-Santini
system (\ref{MSeq}), while the matrix part gives
a matrix system on the background of the Manakov-Santini system
\beaa
\begin{aligned}
&
A_y - B_x=0,
\\
&
(\p_y+v_x\p_x)B-(\p_t+(v_y-u)\p_x) A + u_x A +[A,B]=0
\end{aligned}
\eeaa
For the potential $K$, $A=K_x$, $B=K_y$
we  have
\bea
QK= [K_x,K_y],
\label{extMS}
\eea
where $Q$ is a linear operator
\beaa
Q=\p_t\p_x-\p_y\p_y - (u-v_y)\p_x\p_x -  v_x\p_x\p_y- u_x\p_x.
\eeaa
In the Abelian case instead of equation (\ref{extMS}) we have linear equation
$Q\kappa=0$.
%%%%%%%%%%%%%%%%%%%%%%%%%%%%%%%%%%%%
%%%%%%%%%%%%%%%%%%%%%%%%%%%%%%%%%%%%%%%%%%%%
\section*{Appendix 2}
%%%%%%%%%%%%%%%%%%%%%%%%%%%%%%%%%%%%%%%%%%
\subsection*{Geometric structures}
%%%%%%%%%%%%%%%%%%%%%%%%%%%%%%%%%%%%%%%%%
%%%%%%%%%%%%%%%%%%%%%%%%%%%%%%%%%%%%%%%%%%%%%
The starting point for the geometric interpretation of the systems considered
in this work are two theorems from the article \cite{DFK15}, we also refer the reader to this work 
for more details.

We recall that a conformal structure $[g]$  is called anti-self-dual (ASD)
if the self-dual  part of the Weyl tensor of any $g\in[g]$ vanishes: $W_+=\frac12(W+*W)=0$.
%\subsection*{Local form of EW geometry and dispersionless
%integrable systems}
\begin{theorem}[Dunajski, Ferapontov and Kruglikov (2014)]
\label{prop_ASD}
There exist local coordinates $(z, w, x, y)$ such 
that any ASD conformal structure
in signature $(2, 2)$  is locally represented by a metric 
\bea
\label{ASDmetric1}
{\textstyle\frac{1}{2}}g=dwdx-dzdy-F_y dw^2-(F_x-G_y)dwdz + G_xdz^2,
\eea
where the functions
$F,\;G:M^4\rightarrow \mathbb{R}$ satisfy a 
coupled system of third-order PDEs,
\bea
&&
\p_x(Q(F))+\p_y(Q(G))=0,
\nn\\
&&
(\p_w+F_y\p_x+G_y\p_y)Q(G)+(\p_z+F_x\p_x+G_x\p_y)Q(F)=0,
\label{sd_3rd}
\eea
where
\[
Q=\p_w\p_x-\p_z\p_y+F_y{\p_x}^2-G_x{\p_y}^2-
(F_x-G_y)\p_x\p_y.
\]
\end{theorem}
\begin{theorem}[Dunajski, Ferapontov and Kruglikov (2014)]
\label{theo_ms}
There exists a local coordinate system $(x, y, t)$ 
on $M^3$ such that
any Lorentzian Einstein-Weyl structure is locally 
of the form
\begin{gather}
\begin{split}
g &= -(dy + v_x dt )^2 +4(dx + (u - v_y ) dt ) dt,
\\
\omega &= v_{xx} dy+(-4u_x + 2v_{xy} +v_xv_{xx})dt,
\end{split}
\label{metricMS}
\end{gather}
where the functions  $u$ and $v$ satisfy 
the Manakov-Santini system
\begin{equation*}
\begin{split}
u_{xt} &= u_{yy}+(uu_x)_x+v_xu_{xy}-u_{xx}v_y,
\\
v_{xt} &= v_{yy}+uv_{xx}+v_xv_{xy}-v_{xx}v_y
\end{split}
%\label{MSeq}
\end{equation*}
\end{theorem}
Thus the background equations used in this work have a clear geometric meaning:
SDCS equations (\ref{SDCS2}) define a general (anti)self-dual conformal structure
in signature $(2, 2)$, and the Manakov-Santini system (\ref{MSeq}) corresponds to a general 
Lorentzian Einstein-Weyl structure.

We would also like to mention that
the interpolating equation (\ref{MSint}) was introduced in \cite{Dun08} as ”the most general
symmetry reduction of the second heavenly equation by a conformal Killing
vector with a null self-dual derivative”.

Let us consider
a gauge potential ${A}$, which is a one-form
taking its values in some (matrix) Lie algebra, and
the two-form
${F}=\d{A}+{A}\wedge{A}$
(gauge field).
Matrix equation (\ref{linSDm}) represents (anti)self-dual
Yang-Mills (SDYM)  equations on the background of conformal structure  (\ref{ASDmetric1}),
\bea
{F}={\pm}{*F}, 
\label{SDYM0}
\eea
taken in a special gauge (see \cite{LVB17} for more details). In the Abelian case equations 
become linear, ${F}=\d{A}$,  $\d{A}={\pm}{*\d{A}}$.

Matrix equation (\ref{extMS}) corresponds to the equation
(taken in a special gauge, see \cite{LVB19} for more details)
\bea
D\Phi+\tfrac{1}{2}\omega \Phi=*{F},
\label{Ward}
\eea
where $D\Phi=\d\Phi+[{A},\Phi]$, $\Phi$ is a function
taking values in the Lie algebra (Higgs field, \cite{Dun}).
This equation is considered on the Einstein-Weyl background (\ref{metricMS});
for Minkowski metric it coincides with the
Yang-Mills-Higgs system
introduced by Ward \cite{Dun}, leading to integrable
chiral model.  In the Abelian case this equation becomes linear,
\beaa
\d\Phi+\tfrac{1}{2}\omega \Phi=*{\d{A}}.
%\label{Ward}
\eeaa
%%%%%%%%%%%%%%%%%%%%%%%%%%%%%%%%%%%%%%%%%%%%
\section*{Appendix 3}
%%%%%%%%%%%%%%%%%%%%%%%%%%%%%%%%%%%%%%%%%%
\subsection*{Adjoint Lax operators and a parametric deformation}
%%%%%%%%%%%%%%%%%%%%%%%%%%%%%%%%%%%%%%%%%
To obtain linearisation operator from the Lax pair and construct
its solutions, we will use 
a parametric deformation of the Lax pair introduced in \cite{LVB10},\cite{LVB11}. This deformation 
is nontrivial only for vector fields with nonzero divergence, and in this case 
we have a freedom to add some  term 
with first derivatives containing a parameter to the Abelian extension operator $Q$.

Let us consider a Lax pair of two vector fields of the form (\ref{Lax0})
\bea
X_1=\p_{t_1}
+\sum_{i=1}^N F_i\p_{x_i} + F_0\p_\lambda,\quad
X_2=\p_{t_2}+ \sum_{i=1}^N G_i\p_{x_i}+G_0\p_\lambda.
\label{vfLax}
\eea
We introduce a basic set of wave functions $\Psi^0,\dots,\Psi^N$, a general wave function
is expressed as $\Psi=F(\Psi^0,\dots,\Psi^N)$, $X_1\Psi=0$, $X_2\Psi=0$.
For the linearly degenerate case the terms with $\p_\lambda$ are absent,
$F_0=G_0=0$, $\Psi^0=\lambda$.

A formally adjoint operator to a vector field reads
\beaa
X^*=-(X +\text{div}X),
\eeaa
where $\text{div}X=\p_\lambda F_0 + \sum_{i=1}^N\p_{x_i} F_i$.
We should emphasize that an adjoint operator is no more a pure vector field, it contains
a term $\text{div}X$ which is a multiplication by the function. In the case of zero divergence
(volume-preserving vector field) this term is equal to zero and the vector field is anti-self-adjoint.
Compatibility conditions for adjoint vector fields 
remain the same. It is interesting to note that $a_1=\text{div}X_1$, 
$a_2=\text{div}X_2$ give a special solution to Abelian extension equation 
(\ref{backgroundA}).

Let us consider linear equations corresponding to adjoint vector fields
\bea
(X_1 +\text{div}X_1)\wt\Psi=0, \quad(X_2 +\text{div}X_2)\wt\Psi=0,
\label{vfAdj}
\eea
where by $\wt\Psi$ we denote a wave function of these equations.
A special solution is given by the Jacobian of the basic
wave functions  for vector fields (\ref{vfLax}) (see \cite{LVB11})
$$
\wt\Psi=J:=\frac{\p(\Psi^0,\Psi^1,\dots,\Psi^N)}{\p(\,\l,~ x_1\,,\dots,x_N~)}.
$$
To construct a general solution, we rewrite equations (\ref{vfAdj}) as nonhomogeneous 
linear equations for $\ln \wt\Psi $,
\beaa
X_1\ln \wt\Psi + \text{div}X_1=0, \quad X_2\ln \wt\Psi + \text{div}X_2=0.
\eeaa
Then, evidently, a general solution reads
\beaa
\ln \wt\Psi= \ln J_0 +  F(\Psi^0,\dots,\Psi^N),
\eeaa
and for equations (\ref{vfAdj})
\beaa
 \wt\Psi= J_0  f(\Psi^0,\dots,\Psi^N).
\eeaa

Morover, it is possible to rewrite linear equations in terms of the function 
$J^\alpha$ and obtain a
parametric deformation of the Lax pair
\beaa
X\ln(J^\alpha) + \alpha\text{div}X=0, 
%\\
%(X + \alpha\text{div}X)(\wt\Psi^\alpha)=0,
\eeaa
for equations (\ref{vfAdj})
\beaa
(X_1 +\alpha\text{div}X_1)\wt\Psi_\alpha=0, 
\quad(X_2 +\alpha\text{div}X_2)\wt\Psi_\alpha=0,
%\label{vfAdj}
\eeaa
a general solution to these linear equations  is  of the form
$J^\alpha  f(\Psi^0,\dots,\Psi^N)$.
Using parametrically deformed linear problems, we can add linear term to the operator $Q$, preserving the solvability.
For conformal self-duality equations  (\ref{SDCS3}) the parametric deformation of the Lax pair reads
\beaa
\begin{aligned}
%\label{new_lax}
{X_{1\alpha}}&=\p_z-\lambda\p_x
+F_x\p_x+G_x\p_y+ f_x\p_\lambda + \alpha(F_x+ G_y)_x,
\\
{X_{2\alpha}}&=
\p_w- \lambda\p_y + F_y\p_x+G_y\p_y+f_y\p_\lambda + \alpha(F_x+ G_y)_y.
\end{aligned}
\eeaa 
Instead of relation (\ref{cross}) we have
\begin{multline*}
\bigl(
(\partial_w+F_y \partial_x + G_y\p_y+\alpha(F_{xy}+ G_{yy}))\partial_x
\\
-
(\partial_z+ F_x \partial_x + G_x\p_y + \alpha(F_{xx}+ G_{xy}))\partial_y 
\bigr)\Psi
=\p_\l(f_x\p_y - f_y\p_x ) \Psi,
%\label{cross3}
\end{multline*}
and the parametric deformation of the operator $Q$ looks like
\beaa
Q_\alpha=Q + \alpha( (F_{xx}+G_{xy})\p_y -  (F_{xy}+G_{yy})\p_x)
\eeaa
For the Manakov-Santini system  the deformed Lax pair reads
\bea
\begin{split}
X_{1 \alpha} &=\partial_y-\l\p_x + v_{x}\partial_x + u_x\p_\l + \alpha  v_{xx} ,
\\
X_{2\alpha} &=\partial_t - \l \p_y + (v_{y}-u)\partial_x + u_y\p_\l + \alpha  v_{xy},
\end{split}
\label{defMS}
\eea
and the parametric deformation of the operator $Q$ is
$$
Q_\alpha=Q+\alpha(v_{xy}\p_x -v_{xx}\p_y).
$$

For the case of interpolating equation (\ref{MSint}) the Jacobian is connected with
the basic wave function $\Psi^0$ having 
an expansion $\lambda + u \lambda^{-1} + \dots$ (corresponds
to the Lax-Sato function $L$ in standard dKP hierarchy notations)
by the relation defining interpolating reduction \cite{LVB11},
$$
J=\exp(a(\Psi^0-\lambda)).
$$
The general wave function for the deformed Lax pair with $\alpha=-1$, which is required to
construct symmetries for the interpolating equation, is 
$$
\wt\Psi_{-1}= \e^{a(\mu - \Psi^0 )}\Psi,
$$
where $\Psi$ is an arbitrary wave function $\Psi=F(\Psi^0,\Psi^1)$ of the initial Lax pair. 
%%%%%%%%%%%%%%%%%%%%%%%%%%%%%%%%%%%%%%%%%%

%%%%%%%%%%%%%%%%%%%%%%%%%%%%%%%%%%%%%%%%%%%%
%\subsection*{Conflict of Interest} 
%The author declares no conflicts of interest.
%%%%%%%%%%%%%%%%%%%%%%%%%%%%%%%%%%%%%%%%%%%%%%%%%%%%%%%%

%%%%%%%%%%%%%%%%%%%%%%%%%%%%%%%%%%%%%%%%%%%%%%%%%%%%
\end{document}